\documentclass[final,5p,roman,twocolumn]{elsarticle}

\usepackage{psfrag,slashed,cancel,array,graphicx,hyperref}
\usepackage{subfigure}
\usepackage[utf8]{inputenc}
\usepackage{amsmath} 
\usepackage{mathtools}
\usepackage{mathrsfs}
\usepackage{multirow}
\usepackage{braket}
\usepackage{booktabs} 
\usepackage[normalem]{ulem}
\hypersetup{pdftitle={},pdfcreator={},linkcolor=[rgb]{0.15,0.35,0.75},colorlinks=true,citecolor=[rgb]{0.675,0,0.2},urlcolor=[rgb]{0.15,0.35,0.65}}

\usepackage{comment}
\usepackage{color} 
\usepackage{graphics}

\usepackage{float}

\newcommand{\smallw}{{\scriptscriptstyle W}}
 
\newcommand{\mw}{m_\smallw} 
\newcommand{\mwsq}{m_\smallw^2} 

\newcommand{\smallz}{{\scriptscriptstyle Z}}
\newcommand{\mz}{m_\smallz} 
\newcommand{\mzsq}{m_\smallz^2} 
 
\newcommand{\oa}{${\cal O}(\alpha)$\,}

\newcommand{\oaa}{${\cal O}(\alpha^2)$\,} 
\newcommand{\oaas}{${\cal O}(\alpha\alpha_s)$\,} 
\newcommand{\oaasas}{${\mathcal O}(\alpha\alpha_s^2)$} 
\newcommand{\oaasasas}{${\mathcal O}(\alpha\alpha_s^3)$} 
\newcommand{\oaaas}{${\mathcal O}(\alpha^2\alpha_s)$}

\newcommand{\sefff}{\sin^2\theta_{eff}^{f}\,}

\newcommand{\swtwo}{\sin^2\theta_W}

\newcommand{\Drho}{\Delta\rho}
\newcommand{\Dr}{\Delta r}
\newcommand{\Dk}{\Delta\kappa}
\newcommand{\msbar}{{ \overline{\mathrm{MS}}}}

\def\asr{\bigg( \frac{\alpha_s}{4 \pi} \bigg)}

\allowdisplaybreaks

\begin{document}

\begin{frontmatter}


\title{Three loop QCD corrections to electroweak radiative parameters}

\author[first,second]{Tanmoy Pati}
\author[first,second]{Narayan Rana}
\author[third]{Alessandro Vicini}

\affiliation[first]{organization={School of Physical Sciences, National Institute of Science Education and Research},  
            city={Jatni},
            postcode={752050}, 
            country={India}}

\affiliation[second]{organization={Homi Bhabha National Institute},
            addressline={Training School Complex,  Anushakti Nagar}, 
            city={Mumbai},
            postcode={400094}, 
            country={India}}

\affiliation[third]{organization={Dipartimento di Fisica ``Aldo Pontremoli''},
            addressline={University of Milano and INFN, Sezione di Milano}, 
            city={Milano},
            postcode={I-20133}, 
            country={Italy}}

\date{\today}


\begin{abstract}
We reevaluate the vacuum polarization functions for electroweak gauge bosons at three loops in QCD, employing state-of-the-art perturbative techniques. We apply these results to determine the ${\mathcal{O}}(\alpha \alpha_s^2)$ corrections to the electroweak radiative parameters $\Drho$, $\Dr$ and $\Dk$.
We improve the accuracy of the calculation at this perturbative order, compared to the existing literature, and present some phenomenological implications of these results.  We find
a shift in the prediction of the $W$ boson mass, significant in view of the FCC precision targets.
We improve the prediction of the $\msbar$ electric charge at $q^2=m_Z^2$ 
with the inclusion of these ${\mathcal{O}}(\alpha \alpha_s^2)$ corrections.
\end{abstract}

\end{frontmatter}



\section{Introduction}
The discovery of the Higgs boson at the Large Hadron Collider (LHC) marked the completion of the Standard Model (SM) particle spectrum. However, the subsequent absence of direct signals for New Physics has pivoted the frontier of particle physics towards precision phenomenology. 
As experimental uncertainties continue to shrink, driven by the High-Luminosity LHC \cite{Azzi:2019yne} and the design phases of future colliders \cite{FCC:2025lpp}, the theoretical predictions must reach a comparable level of precision in order to allow a statistically meaningful comparison with the data~\cite{Freitas:2019bre}.

A precision test program requires high-precision theoretical predictions for the cross sections of key scattering processes, where any discrepancy with the data may serve as a signature of physics beyond the SM. At both the LHC and future lepton colliders, this necessitates the simultaneous inclusion of higher-order Quantum Chromodynamics (QCD) and electroweak (EW) corrections,
accounting for their non-trivial interplay.
The Drell-Yan process plays a central role in this program. It has been calculated to next-to-next-to-next-to-leading order (N$^3$LO) in QCD~\cite{Duhr:2020seh,Duhr:2020sdp,Duhr:2021vwj}
and to next-to-next-to-leading order (NNLO) in mixed QCD-EW accuracy
\cite{Bonciani:2021zzf,Buccioni:2022kgy,Dittmaier:2024row,Armadillo:2024ncf}.
The latter contributions turned out to be notably larger than initially anticipated across various kinematic regions, pushing the efforts towards the calculations of complete NNLO EW corrections~\cite{Armadillo:2025mfx,Freitas:2025vax} and 
three-loop mixed QCD-EW corrections~\cite{Pati:2025ivg,Pati:2025xht}.

Beyond scattering cross sections, the SM predicts specific relations among its fundamental parameters, facilitating another suite of high-precision tests.
The most relevant examples are the $W$ boson mass ($\mw$) and the weak mixing angle ($\swtwo$).
In the case of $\mw$, the expression of the muon-decay amplitude in the Fermi theory 
has been matched to the SM expression
incorporating full one-loop~\cite{Sirlin:1980nh} and two-loop \cite{Freitas:2000gg,Awramik:2002wn,Awramik:2002vu,Onishchenko:2002ve,Awramik:2003ee,Awramik:2003rn}
EW corrections. Higher-order QCD effects have been evaluated at
\oaas~\cite{Djouadi:1987di,Kniehl:1988ie,Kniehl:1989yc,Kniehl:1991gu,Djouadi:1993ss}, 
at \oaasas~\cite{Avdeev:1994db,Chetyrkin:1995ix,Chetyrkin:1995js,vanderBij:2000cg,Faisst:2003px},
and at \oaasasas~\cite{Schroder:2005db,Chetyrkin:2006bj,Boughezal:2006xk} for the $\rho$ parameter.
Corrections to the muon-decay amplitude at \oaaas\,
have been presented in refs.~\cite{Chen:2020xot,Dubovyk:2026nhx}.
Regarding the weak mixing angle, the complete set of \oaa corrections to the pseudo-observables used to extract the effective angle at the $Z$ resonance have been presented in ref.~\cite{Dubovyk:2018rlg}, supplemented by the aforementioned QCD corrections to the $\rho$ parameter. 
Current estimates of residual theoretical uncertainties for the $W$ boson mass, derived from global EW fits that account for missing higher orders and parametric uncertainties, are approximately $\pm 6$ MeV \cite{ParticleDataGroup:2024cfk}. Similarly, predictions for the effective leptonic weak mixing angle are subject to a residual uncertainty of $\pm 4.5 \times 10^{-5}$ \cite{deBlas:2025gyz}.

The extraction of these parameters from experimental data relies on a direct comparison with theoretical cross sections, which serve as the primary templates for fitting. In this framework, theoretical uncertainties in the cross sections propagate as systematic errors in the determination of the parameters of interest. These errors can be reduced by incorporating higher-order corrections and refining the parameterization of the proton structure \cite{CarloniCalame:2016ouw,Bagnaschi:2019mzi,Rottoli:2023xdc}.
The current world average for the $\mw$ value is reported with a total uncertainty of 13.3 MeV \cite{LHC-TeVMWWorkingGroup:2023zkn,ParticleDataGroup:2024cfk}. Projections for the High-Luminosity phase of the LHC aim to reduce this error to the 5 MeV level. Furthermore, feasibility studies for an electron-positron collider operating at the $WW$ threshold suggest that a total uncertainty well below 1 MeV is achievable, with projected statistical and systematic contributions of 0.18 MeV and 0.16 MeV, respectively \cite{FCC:2025lpp}.
Similarly, the determination of the effective weak mixing angle at the $Z$ resonance remains dominated by the LEP and SLD measurements, which feature a total error of  $16 \times 10^{-5}$ \cite{ALEPH:2005ab}. 
While the combined results from hadron colliders have become competitive, 
reaching an error of $13 \times 10^{-5}$ \cite{ParticleDataGroup:2024cfk}, 
this level of precision requires a rigorous scrutiny of various theoretical uncertainty sources. 
Looking ahead, a future electron-positron collider could improve the determination of the 
effective weak mixing angle by nearly two orders of magnitude, 
with the final uncertainty estimated to be in the $0.6 \times 10^{-5}$ range \cite{FCC:2018byv}.

These very promising experimental prospects must be matched by an improvement in the precision of the theoretical predictions. Achieving this requires the evaluation of higher-order QCD and EW corrections and a parallel reduction in the uncertainties associated with the key input parameters.
In a recent study \cite{Dubovyk:2026nhx}, the authors investigated 
the ${\cal O}(\alpha^2 \alpha_s)$ corrections to $\Delta r$ \cite{Sirlin:1980nh} and their subsequent impact on the SM prediction for $\mw$. The goal of this paper is to revisit, with state-of-the-art computational techniques, the evaluation of the \oaasas\, corrections to the parameters $\Drho$ \cite{Marciano:1980pb}, $\Dr$, 
$\Dk$ \cite{Marciano:1980pb}, which serve as fundamental building blocks of the EW radiative corrections and are essential for the precise prediction of $\mw$ and $\swtwo$.
The main outcome of this study is that the increased accuracy of our expressions has an impact on the size of the corrections, relevant in view of the future experimental precision target.
We evaluate the self-energy functions via series expansions including more terms compared to the existing literature and we include the massless contributions at third order which were previously missing.
In this paper we also discuss the improvement of the prediction of the $\msbar$-renormalized electric charge, by including the \oaasas\, corrections.

The remainder of this paper is organized as follows: 
Section \ref{sec:theory} defines the relevant EW parameters. 
Section \ref{sec:comp} details our computational procedure and 
Section \ref{sec:result} presents our numerical findings and discussion.

\section{Theoretical Framework}
\label{sec:theory}

\subsection{The $\msbar$-renormalized electric charge}
The electric charge $e$ is defined via Thomson scattering in the limit of zero momentum transfer. Its on-shell (OS) renormalization is achieved by imposing that the counterterm $\delta e = e_0 - e$, with $e_0$ the bare charge, preserves to all orders the value of the renormalized charge, given by the fine structure constant $\alpha(0)=e^2/(4\pi) = 1/137.035999084(21)$ \cite{ParticleDataGroup:2024cfk}.
The electric charge renormalization at two-loop EW level has been presented in ref.~\cite{Degrassi:2003rw}. The proof of the all-order structure of the counterterms has been discussed in ref.~\cite{Dittmaier:2021loa}.
Due to the validity of the QED Ward identities, the OS electric charge counterterm depends only on the photon vacuum polarization,
making manifest the universality of the electric charge.
Within the SM, the validity of QED-like Ward identities can be maintained to all orders by adopting a gauge fixing chosen according to the Background Field Method \cite{Denner:1994xt,Degrassi:2003rw,Dittmaier:2021loa}.
This approach allows the renormalization counterterm to be defined through the photon self-energy only.

The transverse part of the photon vacuum polarization can be written as
$\Pi^T_{\gamma\gamma}(s)=e_0^2\, s\,\Pi_{\gamma\gamma}(s)$, and the function $\Pi_{\gamma\gamma}(s)$
can be decomposed into several distinct contributions according to the particles coupling to the external photons: leptons ($\ell$), bosons ($b$), light quarks ($5$), and perturbative contributions from the top quark ($p$).
To address the non-perturbative nature of the light-quark contributions at $s=0$,
$\Pi_{\gamma\gamma}^{(5)}(0)$ is replaced by an experimental quantity, 
$\Delta\alpha_{had}^{(5)}(\mzsq)$,
linked to the photon vacuum polarization and its perturbative remainder via a dispersion relation.
\begin{align}
 \Pi_{\gamma\gamma}(0) &= 
 \Pi_{\gamma\gamma}^{(\ell)}(0)+
 \Pi_{\gamma\gamma}^{(b)}(0)+
 \Pi_{\gamma\gamma}^{(p)}(0)+
 \Pi_{\gamma\gamma}^{(5)}(0)   
 \nonumber\\ &
 = \Pi_{\gamma\gamma}^{(\ell)}(0) 
 + \Pi_{\gamma\gamma}^{(b)}(0) 
 + \Pi_{\gamma\gamma}^{(p)}(0)
 + \mathrm{Re}\Pi_{\gamma\gamma}^{(5)}(\mzsq)
 \nonumber\\&
 ~~~~ + \mathrm{Re}\left[ \Pi_{\gamma\gamma}^{(5)}(0)- \Pi_{\gamma\gamma}^{(5)}(\mzsq) \right] ,
 \label{eq:Pigammagammazero}
\end{align}
where $\Delta\alpha_{had}^{(5)}(\mzsq) = 4 \pi \alpha \, \mathrm{Re} [ \Pi_{\gamma\gamma}^{(5)}(0)- \Pi_{\gamma\gamma}^{(5)}(\mzsq) ]$
includes all the non-perturbative corrections due to the strong interaction at low momenta.
The term $\mathrm{Re}\, \Pi_{\gamma\gamma}^{(5)}(\mzsq)$ can be safely evaluated with a perturbative approach.

The electromagnetic coupling can be defined within the $\msbar$ renormalization scheme, as
\begin{equation}
 \alpha_{\msbar}(\mz^2)=\frac{\alpha(0)}{1 - \Delta\alpha_\msbar (\mz^2)} \,.
\end{equation}
$\Delta \alpha_\msbar (\mz^2)$ is given by the finite part of the counterterm, subtracted of its ultraviolet (UV) pole according to the $\msbar$ prescription and is evaluated with the renormalization scale $\mu_R=\mz$.
\begin{align}
\label{eq:alphamsbar}
\Delta\alpha_\msbar(\mzsq) &= - 4 \pi \alpha
    \left[ \hat\Pi_{\gamma\gamma}^{(\ell)}(0)
         + \hat\Pi_{\gamma\gamma}^{(b)}(0)
         + \hat\Pi_{\gamma\gamma}^{(p)}(0)
    \right]
\nonumber\\
    &~~~ - 4 \pi \alpha\, \Pi_{\gamma\gamma}^{(5)}(\mzsq) 
      - \Delta\alpha_{had}^{(5)}(\mzsq) \,.
\end{align}
The symbols $\hat\Pi_{\gamma\gamma}(s)$ represent the finite part of the photon vacuum polarization, after removing the UV poles according to the $\msbar$ prescription.
In this paper we compute the \oaasas\, contribution
to $\Delta \alpha_\msbar (\mz^2)$.
Only $\hat\Pi_{\gamma\gamma}^{(p)}(0)$ and $\Pi_{\gamma\gamma}^{(5)}(\mzsq)$ contribute at \oaas and \oaasas.
We obtain an accurate prediction of $\alpha_\msbar(\mz^2)$, which can be used in turn as boundary condition for the evaluation of the coupling at arbitrary scales $\mu$, solving the Renormalization Group Equation.

\subsection{The electroweak parameters $\Drho,\Dr,\Dk$ }
The strength of the weak charged-current interaction is parameterized by the Fermi constant, 
$G_\mu$, which is determined with high precision from measurements of the muon lifetime within the Fermi theory. The current experimental value is
$G_\mu = 1.11663788(6) \times 10^{-5} $ GeV$^{-2}$ \cite{ParticleDataGroup:2024cfk}.
By equating the muon-decay amplitudes at zero momentum transfer in both the Fermi theory and the SM, 
one establishes a fundamental relation between $G_\mu$, $\alpha (0)$, $\mw$ and $\mz$.
\begin{equation}
    G_\mu= \frac{\pi\, \alpha(0)}{\sqrt{2} (1-\mwsq/\mzsq)~  \mwsq}(1+\Delta r) \,.
    \label{eq:Deltar}
\end{equation}
The parameter $\Dr$ \cite{Sirlin:1980nh} is a finite physical quantity accounting for radiative corrections to the muon-decay amplitude, 
after subtracting the QED components that are identical in both the Fermi theory and the SM. 
This relation defines a renormalized coupling within the SM that characterizes the strength of the weak charged-current interaction. 
Solving Eq.~(\ref{eq:Deltar}) for $\mw$ provides a theoretical prediction for 
the $W$ boson mass. The impact of higher-order radiative corrections on $\Dr$ and 
the resulting $\mw$ prediction have been discussed extensively 
in refs.~\cite{Freitas:2000gg, Awramik:2003rn, Degrassi:2014sxa, Freitas:2025vax}.

At \oa\,the radiative corrections to the muon-decay amplitude include 
self-energy contributions ($\Pi_{WW}$) to the $W$ propagator, 
vertex ($V_W$) and box ($B_W$) corrections and the renormalization of 
the tree-level amplitude parameters.
Working within the OS scheme, we define the gauge-boson mass counterterms from the transverse part of the corresponding self-energies: $\delta m_V^2 = \mathrm{Re}\left( \Pi^T_{VV} (m_V^2) \right)$.
In this framework, $\Dr$ can be explicitly expressed as:
\begin{align}
 \Delta r &= \frac{1}{m_W^2} \left[ \Pi^T_{WW}(0) - \text{Re}( \Pi^T_{WW}(m_W^2) ) \right]  
 + V_W + B_W 
 \nonumber\\
 &+ 2 \frac{\delta e}{e}
  + \frac{c_w^2}{s_w^2}\text{Re} \left(
      \frac{\Pi^T_{WW}(m_W^2)}{\mwsq} 
    - \frac{\Pi^T_{ZZ}(\mzsq)}{\mzsq}
    \right) \,,
\end{align}
where $s_w^2=1-\mwsq/\mzsq$ is the squared sine of the weak mixing angle and $c_w^2=1-s_w^2$.
Higher-order QCD corrections, e.g. at \oaas,\, and \oaasas\,, enter $\Dr$ exclusively through 
the quark loop contributions to the gauge-boson self-energies. 
The vertex ($V_W$) and box ($B_W$) terms do not receive corrections at these perturbative orders.

The $\rho$ parameter quantifies the relative strength of the neutral-current weak coupling 
compared to the charged-current coupling. At the tree level, this quantity is exactly one, because of the accidental global custodial symmetry of the SM. The radiative corrections break the symmetry, via the mass splitting in the fermion doublets and via hypercharge corrections, yielding $\Drho=\rho-1\neq 0$. The largest effects stem from the third family, because of the large difference between the top and bottom quark masses.
The universal, i.e. process independent, contributions to $\Drho$ can be identified in the comparison of neutral- and charged-current elastic lepton-neutrino scattering \cite{Marciano:1980pb}:
\begin{equation}
    \Drho(0) = 
    \frac{\Pi_{ZZ}^T(0)}{\mzsq}
    -
    \frac{\Pi_{WW}^T(0)}{\mwsq}\,.
\end{equation}
As a remnant of the underlying custodial symmetry, $\Drho(0)$ is UV-finite at the one-loop level. 
It does not require independent renormalization at its calculation order, as it only depends on the renormalization of lower-order amplitudes. This makes $\Drho(0)$ a highly sensitive and robust prediction for testing the SM at the precision frontier.
The QCD corrections to $\Drho(0)$ are relevant only for the top-bottom doublet case and have been computed up to the four-loop level
\cite{Djouadi:1987di,Kniehl:1988ie,Kniehl:1989yc,Kniehl:1991gu,Djouadi:1993ss,Avdeev:1994db,Chetyrkin:1995ix,Chetyrkin:1995js,vanderBij:2000cg,Faisst:2003px,Dubovyk:2026nhx,Schroder:2005db,Chetyrkin:2006bj,Boughezal:2006xk}.
If we assume that the light quarks of the first two families are all massless, then their contribution to $\Drho(0)$ vanishes.
It is possible to recast the formula for $\Dr$ as:
\begin{equation}
 \Dr = \Delta\alpha(\mzsq) - \frac{c_w^2}{s_w^2} \Drho(0) + \Delta r_{rem}.
 \label{eq:Deltarcontributions}
\end{equation}
The first term is responsible for the large ($\sim$ 7\%) running of the electric charge.
$\Drho(0)$ contains the large heavy fermion effects ($\sim$ -3.5\%, due to the enhancement by the factor $c_w^2/s_w^2\simeq 3.5$) 
of the $\rho$ parameter, while the last term is a finite contribution specific of the muon-decay process.

A third set of EW radiative corrections is responsible for the redefinition of the weak mixing angle at the quantum level. Such effects can be studied from the analysis of the form factor describing the coupling of a $Z$ boson to a fermion $f$.
The strength of the V--A component of the neutral-current interaction, expressed in units of $G_\mu$, 
is parameterized by a factor $\rho^f(q^2)$ \cite{Degrassi:1990ec}.
This factor includes fermion-specific and $q^2$-dependent corrections beyond those present in $\Drho(0)$.
Any residual deviation between this coupling and the electromagnetic component of the $Z$ current
is parameterized by the factor $\kappa (q^2) \equiv 1 + \Delta\kappa (q^2)$.
Under this construction, the $Z f{\bar f}$ vertex function reads
\begin{align}
 \Gamma_{Zf\bar f}^\mu(q^2) &= i \bar{z}_f  \gamma^\mu
                       \bigg( T_3^f \frac{1-\gamma_5}{2} - \kappa(q^2) \sin^2\theta_W\, Q_f \bigg),
\label{eq:Zvertexrad}
\end{align}
where $\bar{z}_f = 2 m_Z (\sqrt{2} G_\mu)^{\frac{1}{2}} \, \rho^f (q^2)$.
In Eq.~(\ref{eq:Zvertexrad}),
the product $\kappa(q^2) \swtwo$ can be interpreted as the square of the sine of a renormalized mixing angle that absorbs the radiative corrections $\Delta\kappa$. 
At the $Z$ resonance, this parameterization is particularly relevant, because the form factor is gauge invariant. This allows for the definition of an effective, fermion species dependent, 
weak mixing angle $\sefff$, which reabsorbs the effect of all those radiative corrections 
that differently affect the left- and the right-handed components of the fermionic current.
The effective weak mixing angle can be determined experimentally 
(cfr. the extraction in terms of pseudo-observables \cite{ALEPH:2005ab} or with 
a direct approach relevant at hadron colliders \cite{Chiesa:2019nqb}) and it can be compared with its theoretical prediction in the SM \cite{Degrassi:1990ec,Degrassi:1996ps},
which can be computed starting either from an OS or from an $\msbar$-renormalized weak mixing angle, 
and evaluating the respective contributions to the $Z$ form factor.
\begin{equation}
    \sin^2\theta_{eff}^f
    =
    \kappa^f(\mzsq)\sin^2\theta_{OS}
    =   \hat\kappa^f(\mzsq)\sin^2\theta_\msbar \,.
    \label{eq:sefflth}
\end{equation}
The corrections to $\Delta \kappa$ stem from the $\gamma-Z$ self-energy contributions, the renormalization of the tree-level mixing angle present in the vector coupling of the $Zf\bar{f}$ vertex function, and the vertex corrections.
The oblique corrections to $\Delta \kappa$ at \oaas and \oaasas\, enter via the quark self-energies as follows:
\begin{align}
 \Delta \kappa^{oblique}_{QCD} &= \frac{c_w^2}{s_w^2} \text{Re} 
                   \bigg( \frac{\Pi_{ZZ}(\mz^2)}{\mz^2} - \frac{\Pi_{WW}(\mw^2)}{\mw^2} \bigg)
\nonumber\\&                   
                  - \frac{c_w}{s_w} \frac{\Pi_{\gamma Z}(\mz^2)}{\mz^2} \,.
\label{eq:deltakappaQCD}                  
\end{align}

\subsection{The finite renormalization combinations $\delta c_V$}
The one-loop EW corrections to the production of vector bosons at hadron colliders necessitate the renormalization of tree-level parameters and the wave-functions (WFs) of external OS particles.
A significant simplification occurs within the framework of the BFG, because the gauge fixing of the background fields is chosen in a way to preserve QED-like Ward Identities (WIs). In the case of $W$ and $Z$ bosons, other WIs relate the WF renormalization constants to the charge counterterms.
The QCD corrections to first and second order in $\alpha_s$, on top of an \oa\,kernel, preserve these relations \cite{Bonciani:2019nuy,Bonciani:2020tvf,Pati:2025xht}, while the same statement does not hold for purely EW corrections, starting from \oaa.
We thus find it convenient to identify specific combinations of charge and WF counterterms which turn out to be UV finite and are building blocks which systematically appear in any higher-order calculation.

The definition of $\delta c_V$ (for $V=A,Z,W$) depends on the input scheme. 
As such, with $(\alpha(0), m_W, m_Z)$ in input, 
they are defined as:
\begin{align}
\delta c_A &= \delta \, Z_{AZ} 
           - \frac{2 c_w s_w}{s_w^2} \frac{\delta c_w^2}{c_w^2} 
           =\frac{2s_w}{c_w}\Delta\kappa_{QCD}^{oblique} \,.    
\nonumber\\
 \delta c_Z &= \delta \, Z_{ZZ} + 2 \,  \frac{\delta e}{e}
            + \frac{c_w^2-s_w^2}{s_w^2} \frac{\delta c_w^2}{c_w^2} \,.    
\nonumber\\
 \delta c_W &= \delta \, Z_{WW} + 2 \,  \frac{\delta e}{e}
            + \frac{c_w^2}{s_w^2} \frac{\delta c_w^2}{c_w^2} \,. 
\end{align}
Here $Z_{VV} = 1 + \delta \, Z_{VV}$ is the WF renormalization constants of the OS vector boson $V$.

\section{Computational procedure}
\label{sec:comp}
We follow an automated computational workflow to evaluate the vacuum polarization
functions for both charged and neutral gauge bosons.
Corresponding Feynman diagrams were generated with \textsc{QGRAF} \cite{Nogueira:1991ex}. 
For the third generation of quarks ($t,b$), we identify 70 diagrams contributing to $\Pi_{ZZ}$ and 31 diagrams for $\Pi_{WW}$.
The raw output from \textsc{QGRAF} was subsequently processed by in-house \textsc{FORM} \cite{Tentyukov:2007mu} routines, 
which translates the diagrams into Feynman amplitudes and performs the Dirac, Lorentz, and color algebra. 
The resulting expressions consist of $\sim {\mathcal{O}(10^4)}$ scalar Feynman integrals. 
We employ the standard method of Integration-By-Parts (IBP) reduction \cite{Tkachov:1981wb,Chetyrkin:1981qh,Laporta:2000dsw}
to reduce the scalar integrals to 94 Master Integrals (MIs). 
We have utilized the public codes \textsc{Kira}~\cite{Maierhofer:2017gsa,Klappert:2020nbg,Lange:2025fba} 
and \textsc{LiteRed}~\cite{Lee:2012cn,Lee:2013mka} to perform the IBP reduction.

The primary task lies in the evaluation of the three-loop two-point MIs at the 
kinematic scales $q^2 = 0$ and $q^2 = m_V^2$ (where $V \in \{Z,W\}$). 
From two-loop onward, the analytic structure of the MIs becomes complex, 
frequently involving elliptic multiple polylogarithms. 
Hence, we solve the MIs using a system of differential equations evaluated via a series expansion. 
Given that the ratio $z_t \equiv m_Z^2/m_t^2 \simeq 0.28$ resides well within the radius of convergence 
for a series centered at zero, the expansion provides sufficient accuracy,
eliminating the need for full closed-form analytic solutions at $q^2 = m_V^2$. 
We determine these series solutions using the generalized Frobenius method, implemented through an in-house \textsc{Mathematica} routine.
The boundary conditions at $q^2\to 0$ are determined through a combination of different techniques: the 
OS-reduction method, the request of regularity of the solution at the pseudo-thresholds, and, for a few cases, 
high-precision PSLQ fits \cite{pslq:92} via \textsc{PolyLogTools}~\cite{Duhr:2019tlz} based on numerical evaluations using \textsc{AMFlow}~\cite{Liu:2022chg}.
In addition to multiple zeta values (MZVs), $\ln(2)$ and $\ln(3)$, 
our results involve the following set of transcendental constants~\cite{broadhurst1992three,Kalmykov_2005,Martin:2016bgz,Schr_der_2005}
and their combinations
\begin{equation*}
    \left\{
    \begin{gathered}
        \sqrt{3} \pi \zeta(2),
        \sqrt{3} \pi \zeta(2) \ln(3),
        \sqrt{3} \pi \zeta(3),
        \sqrt{3} \, \text{Cl}_2 (\psi),
        \\
        \sqrt{3} \, \text{Cl}_2 (\psi) \ln(3),
        \sqrt{3} \, \text{Cl}_2 (\psi) \ln^2(3),
        \sqrt{3} \, \text{Cl}_2 (\psi) \zeta_2,
        \\
        \text{Cl}^2_2 (\psi),
        \text{Li}_4 (1/2),
        \sqrt{3} \, \text{Ls}_3 (2 \psi),
        \\
        \sqrt{3} \, \text{Ls}_3 (2 \psi) \ln(3),
        \sqrt{3} \, \text{Ls}_4 (\psi),
        \sqrt{3} \, \text{Ls}_4 (2 \psi)
    \end{gathered}
    \right\}
\end{equation*}
where $\psi = \pi/3$. 
Clausen functions and log-sine integrals are defined as 
\begin{align}
 & \text{Cl}_2 \left( \phi \right) = \text{Im} \left[ \text{Li}_2 \left( e^{i \phi} \right) \right] \,.
 \\
 & \text{Ls}_n \left( \phi \right) = \int_{0}^{\phi} \ln^{n-1} \left| 2 \sin \frac{x}{2} \right| dx \,.
\end{align}
While all these constant appear in the intermediate stages, they cancel out in the final expressions, except for MZVs and $\text{Cl}_2 (\psi)$.

We have performed the UV renormalization using a mixed scheme: the heavy quark mass is 
renormalized within the OS framework, while $\alpha_s$ is treated in the $\overline{\text{MS}}$ scheme. 
The required mass-renormalization constants are adopted from refs.~\cite{Broadhurst:1991fy,Melnikov:2000zc,Marquard:2007uj}.

\section{Results}
\label{sec:result}
In this section, we present the ${\mathcal{O}}(\alpha \alpha_s^2)$ corrections to
the electric charge renormalization,
$\Delta \rho (0)$, $\Delta r$ and $\Delta \kappa$.
We also discuss ${\mathcal{O}}(\alpha \alpha_s^2)$ contributions to the particular 
UV finite renormalization combinations $\delta c_V$. 
Results for all these parameters discussed in this work are provided in an ancillary file 
attached to the \textsc{arXiv} submission.
At \oaasas\, we provide a fully independent re-evaluation and validation of the existing three-loop results for $\Drho(0)$, $\Delta r$ and $\Delta \kappa$~\cite{Chetyrkin:1995js}. We further improve $\Dr$ and $\Delta\kappa$ by adding the three-loop light-quark contributions. In the evaluation of the self-energy functions by series expansion, we include a larger number of terms compared to those considered in the seminal work~\cite{Chetyrkin:1995js}, and check the corresponding numerical impact.
We observe that this improved theoretical accuracy has an impact on the predictions, not negligible in view of the precision target at the FCC.
In the numerical evaluation, we adopt the following input parameters:
\begin{equation}
\begin{gathered}
 m_Z = 91.1876, \;
 m_W = 80.385,\; m_t = 173.2, \; m_H = 125.25,
 \nonumber\\
 \alpha(0) = 0.00729735, \,
 \alpha_s(m_Z)=0.1178, \,
 \alpha_s(m_t)=0.1075.
\end{gathered}
\end{equation}
To address the ongoing discrepancy between lattice QCD data and low-energy dispersive results, 
we use for $\left.\Delta \alpha\right|^5_{had}(\mzsq)$ a mean value derived from both approaches, incorporating the recent estimate for 
hadronic vacuum polarization up to the $Z$ scale, 
$\left.\Delta \alpha\right|^5_{had}(\mzsq)=(2780.7\pm1.1)\times 10^{-5}$ \cite{Erler:2023hyi}.

\subsection{The electric charge in the $\msbar$ scheme}
\noindent
We compute the electric charge counterterm at \oaasas\, in the $\overline{\text{MS}}$ scheme.
We define the QED $\beta$-function as
\begin{equation}
    \beta_{QED}=- \sum_{i,j=0}^\infty
    \alpha_s^i \alpha^{j+1} \beta^{(i,j)}_{e} \,.
\end{equation}
Our result for the two-loop QCD corrections to $\beta_{QED}$ agrees with the known result \cite{Billis:2019evv}:
\begin{equation}
    \beta^{(2,0)}_e = N \sum_q e_q^2\left(2 C_F^2 -\frac{133}{9} C_F C_A  + \frac{22}{9} C_F n_F \right),
\end{equation}
where $C_A = N$ and $C_F = (N^2-1)/(2 N)$ are the Casimirs of the SU(N) gauge theory, and $n_F=6$ denotes the number of quark flavors. 
Furthermore, we have verified the \oa and \oaa fermionic contributions to 
$\beta^{(0,1)}$ and $\beta^{(0,2)}$ respectively, in the $\overline{\text{MS}}$ scheme, against the results of ref.~\cite{Degrassi:2003rw}, finding perfect agreement. 
Table~\ref{tab:alpha} presents the perturbative contributions from both heavy ($\Pi_{\gamma\gamma}^{(p)}(0)$) and light ($\Pi_{\gamma\gamma}^{(5)}(m_Z^2)$)  quarks 
to the shift $\Delta \alpha_\msbar (m_Z^2)$, as defined in Eq.~(\ref{eq:alphamsbar}),
in units of $10^{-5}$ at the renormalization scale $\mu=\mz$. 
\begin{table}[H]
\centering
\setlength{\tabcolsep}{5pt}
\renewcommand{\arraystretch}{1.5}
  \begin{tabular}{|c|c|c|c|}
  \hline
    Order & \oa &   \oaas &   \oaasas \\
  \hline\hline
   $\Delta \alpha (m_Z^2) |_{q}$ & 340.71 & 7.16 & -1.88 \\
  \hline
  \end{tabular}
  \caption{Total perturbative contributions from heavy and light quarks to $\Delta \alpha_{\msbar} (m_Z^2)$ at different perturbative orders.}
  \label{tab:alpha}
\end{table}
By incorporating these three-loop QCD effects along with complete one- and two-loop EW and QCD corrections \cite{Degrassi:2003rw}, 
we achieve a more precise determination of $\alpha_\msbar (m_Z^2)$:
\begin{equation*}
 \alpha^{-1}(m_Z^2) = 128.079  \pm 0.015 \,.
\end{equation*}
%

\subsection{\texorpdfstring{$\rho$}{rho} parameter}
\noindent
The factor $\Drho(0)$ admits a perturbative expansion in $\alpha_s$, which reads, assuming a massless bottom quark:
\begin{align}
 \Delta \rho(0) = \frac{3 G_F m_t^2}{8 \sqrt{2} \pi^2} 
   \bigg\{ 1 +  
  \sum_{m=1}^{\infty} \asr^m \delta \rho^{(m)}  \bigg\} \,.
\end{align}
$\delta \rho^{(m)}$ for $m=1, 2$, have been presented in \cite{Chetyrkin:1995ix}.
We recompute $\delta \rho^{(1)}$ and $\delta \rho^{(2)}$ as the following combinations of transcendental constants: 
\begin{align}
 \delta \rho^{(1)} &= C_F \bigg(-2-4\zeta_2 \bigg),
\\
 \delta \rho^{(2)} &= 
  C_F^2 \bigg( 
-\frac{238}{9}
-11 \zeta_4
-\frac{770}{9} \zeta_2
+\frac{1084}{3} \zeta_3
+2 c_4
 \bigg)
%
\nonumber\\&
 + C_A C_F \bigg(
-\frac{49}{6}
+\frac{75}{2} \zeta_4
-\frac{98}{3} \zeta_2
-\frac{452}{3} \zeta_3
-c_4
 \bigg)
\nonumber\\&
 + C_F \bigg( 
 \frac{188}{3}
+108 c_2
-\frac{100}{3} \zeta_2
-\frac{512}{3} \zeta_3
 \bigg)
\nonumber\\&
 + C_F n_F \bigg(
-\frac{2}{3}
+\frac{52}{3} \zeta_2
-\frac{16}{3} \zeta_3
 \bigg) \,,
\end{align}
where, $c_2 = 2 ~ \text{Cl}_2 (\psi)/\sqrt{3}$ with $\psi = \pi/3$ and
\begin{align}
 c_4 &= -12   \left( \text{Cl}_2 \left( \psi \right) \right)^2
+64~ \text{Li}_4 \left(1/2\right)
-78 \sqrt{3}~ \text{Cl}_2 \left( \psi \right) 
\nonumber\\&
-\frac{52}{5} \zeta_2^2
+48 \zeta_2 \ln(2)
-16 \zeta_2 \ln^2(2)
+\frac{8}{3}  \ln^4(2) \,.
\end{align}
At \oaasas\, only a numerical value for the $c_4$ constant was previously known.
%


\subsection{Numerical results for $\Dr$ and $\Delta \kappa$}
\begin{figure}[H]
 \centering
 \includegraphics[width=0.42\textwidth]{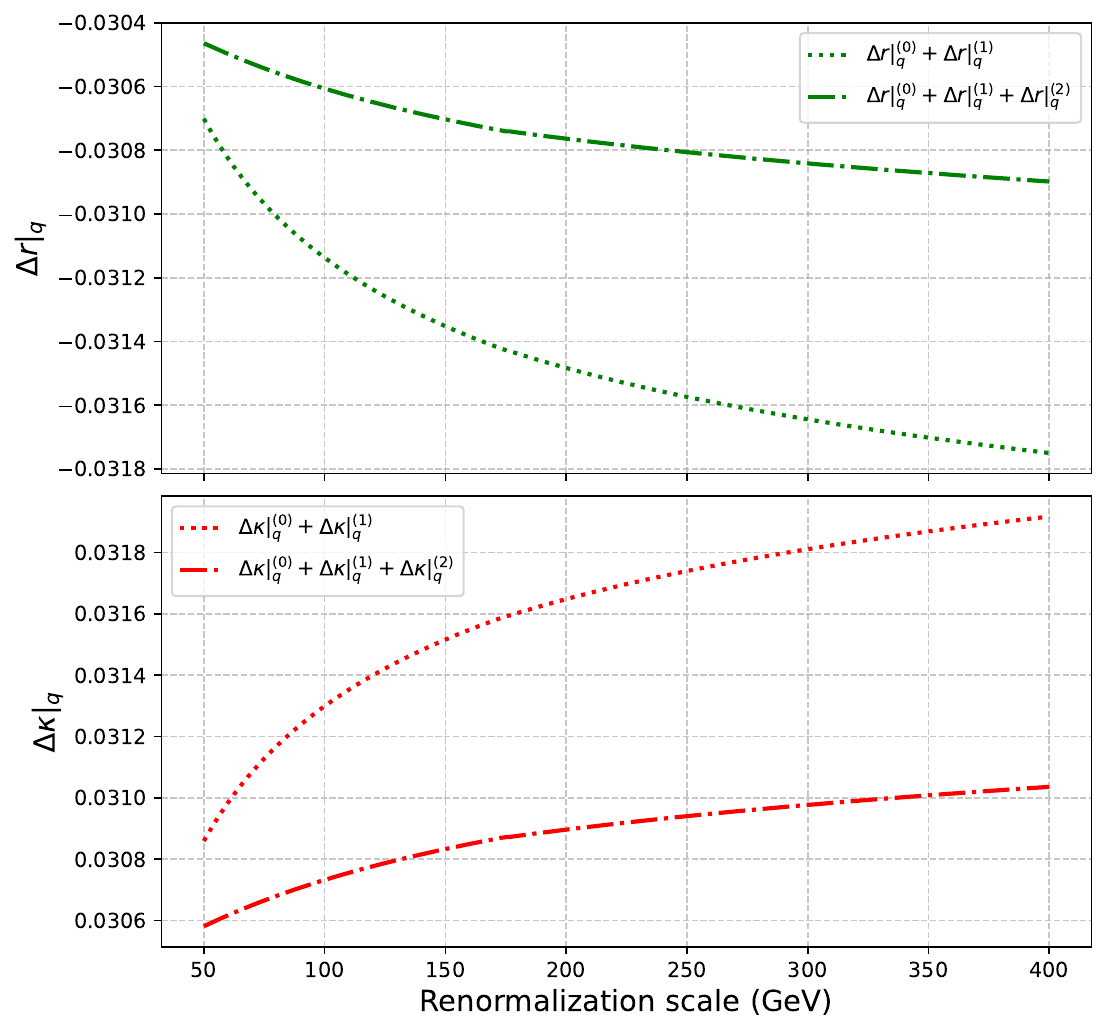}
 \caption{Renormalization scale variation of $\Delta r|_q$ and $\Delta \kappa|_q$
 including contributions at different orders.}
 \label{fig:kappa}
\end{figure}

\begin{figure}[H]
 \centering
 \includegraphics[width=0.42\textwidth]{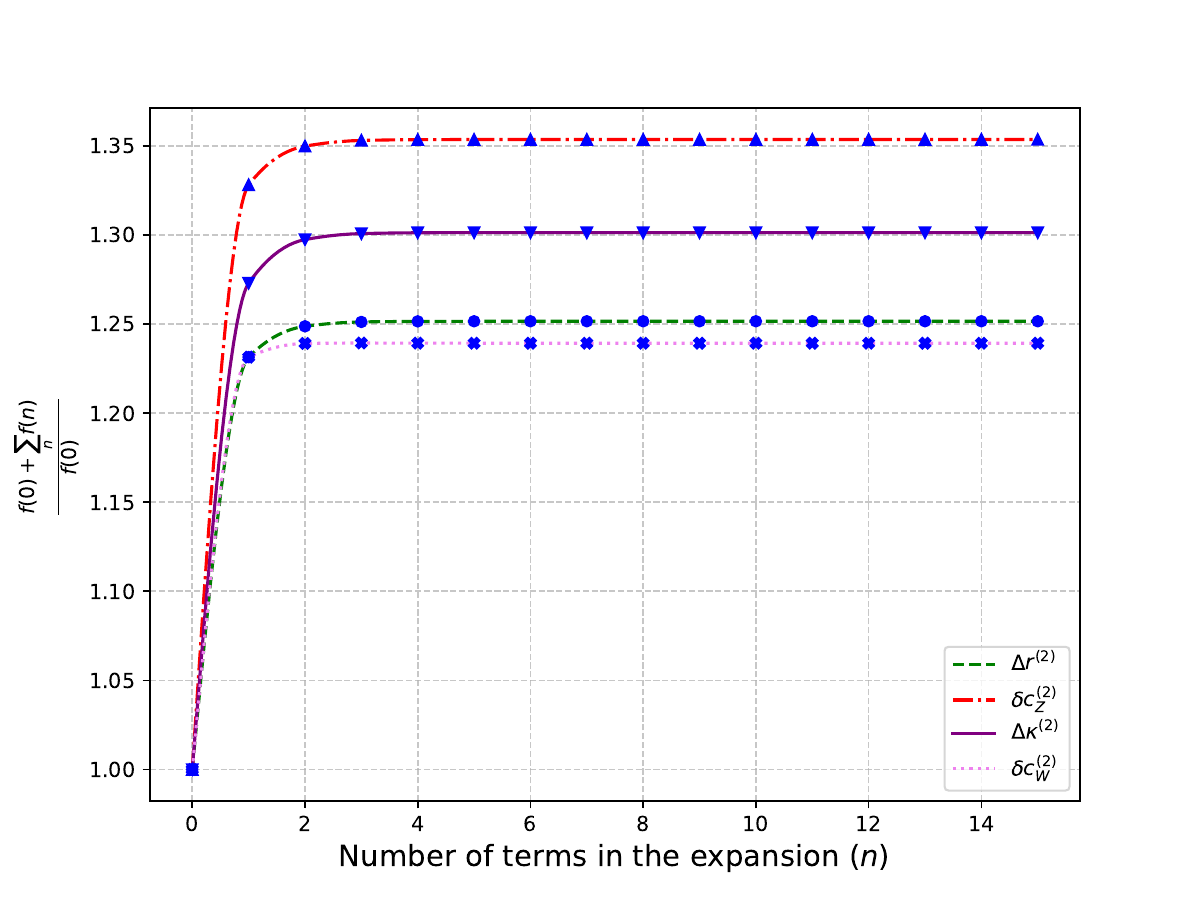}
 \caption{Convergence of the series expansion used to compute EW parameters at \oaasas.}
 \label{fig:seriesConv}
\end{figure}

\noindent
The QCD corrections to the EW first order expressions of the  
parameters $\Delta r$ and $\Delta \kappa$ affect only the quark self-energies. They 
admit a perturbative expansion in $\alpha_s$ as:
\begin{align}
 \Delta r|_q &= - \frac{3 G_F m_t^2}{8 \sqrt{2} \pi^2} \, \frac{c_w^2}{s_w^2} \, \bigg\{
  \sum_{m=0}^{\infty} \asr^m \Delta r^{(m)}  \bigg\} \,.\\
 \Delta \kappa|_q &= - \frac{3 G_F m_t^2}{8 \sqrt{2} \pi^2} \, \frac{c_w^2}{s_w^2} \, \bigg\{
  \sum_{m=0}^{\infty} \asr^m \Delta \kappa^{(m)}  \bigg\} \,.
\end{align}
We have computed the three-loop QCD corrections $\Delta r^{(2)}$ and $\Delta \kappa^{(2)}$,
as a Taylor series in $z_t$ up to ${\mathcal{O}}(z_t^{15})$. 
Eq.~(\ref{eq:Deltarcontributions}), 
the $\Delta \alpha(m_Z^2)$ contribution to $\Delta r$, includes the photon self-energy 
$\hat{\Pi}_{\gamma\gamma}^{(p)}$ which we evaluate at $q^2 \rightarrow 0$,
to ensure consistency with Eq.~(\ref{eq:Pigammagammazero}). 
We note that, in ref.~\cite{Chetyrkin:1995js}, 
$\hat{\Pi}_{\gamma\gamma}^{(p)}$ had been evaluated at $q^2 = m_Z^2$.
We find full agreement with their results after accounting for appropriate $\hat{\Pi}_{\gamma\gamma}^{(p)}$.
The full analytical expressions are provided in the ancillary files accompanying this manuscript.
In the following, we present the numerical evaluations of the coefficients $\Delta r^{(m)}$ and $\Delta \kappa^{(m)}$, with $n_F=6$ flavors and $\mu^2 = m_t^2$.
Each correction is split in the contribution of 
the $(t,b)$ doublet and that of the light quark flavors $(u,d,c,s)$.
In Table \ref{tab:kZW} we show the values of $\Delta r^{(m)}$ and $\Delta \kappa^{(m)}$ for $m=0,1,2$.
\begin{table}[H]
\centering
\begin{tabular}{|c r|r|r|}
\hline
 $m$ & & $\text{Re}(\Delta r^{(m)})$ & $\text{Re}(\Delta \kappa^{(m)})$   \\
\hline\hline
\multirow{2}{*}{0}   & {\footnotesize ${tb}$}  & 1.2083    &  $-1.2560$ \\
                     & {\footnotesize ${lq}$}  & -0.1033   &  $0.1449$ \\
\hline
\multirow{2}{*}{1}   & {\footnotesize ${tb}$}  & -12.6775  &  $12.6269$ \\
                     & {\footnotesize ${lq}$}  & -0.4132   & $0.5794$  \\
\hline
\multirow{2}{*}{2}   & {\footnotesize ${tb}$}  & -292.2481 & $303.8672$  \\
                     & {\footnotesize ${lq}$}  & -6.7453   & $9.4582$  \\
\hline
\end{tabular}
\caption{Contributions to $\Dr|_q$ and $\Delta \kappa|_q$, at different orders in QCD.}
\label{tab:kZW}
\end{table}
We find that the dependence of $\Dr|_q$ and $\Delta\kappa|_q$ on the QCD renormalization scale $\mu_R$ is reduced upon inclusion of the \oaasas\, corrections, as shown in Fig.~\ref{fig:kappa}.
Ref.~\cite{Chetyrkin:1995js} provides analytic results for $\Dr$ and $\Dk$ up to 
$\mathcal{O}(z_t)$, alongside a numerical evaluation extending to $\mathcal{O}(z_t^2)$.
Our results include all the terms up to $z_t^{15}$.
As shown in Fig.~\ref{fig:seriesConv},
the series exhibits moderate fluctuations at $\mathcal{O}(z_t)$
but reaches stability beyond $\mathcal{O}(z_t^3)$.
We also evaluated the Taylor series for the MIs around $q^2 = m_V^2$ using \textsc{AMFlow} 
for the numerical boundary values. The results agree with the Taylor expansion around $q^2 = 0$ to 16 digits for all parameters.

\subsection{Effects on $\mw$ and $\sefff$}
\noindent
We provide an updated evaluation of the $W$-boson mass shift by incorporating the improved \oaasas\,
contributions computed in this work. In the OS scheme, the $W$-boson mass is expressed \cite{Consoli:1989fg} as:
\begin{equation}
 m_W^2=\frac{m_Z^2}{2} \Bigg (1 + \sqrt{1-\frac{2\sqrt{2}\pi \alpha}{G_F m_Z^2}(1+\Delta r)} \,\, \Bigg) \,.
\end{equation}
By employing the precisely measured experimental values of $\alpha$, $G_\mu$ and $\mz$, 
the value of $\mw$ can be determined from the above relation. 
Radiative corrections, encapsulated in $\Delta r$, induce a shift in the $W$ mass from its tree-level value. 
We denote this shift as $\delta m_W$ and summarize the two- and three-loop fermionic contributions from both the heavy ($tb$) and light ($lq$) quark flavors in Table~\ref{tab:Wmass}.
It is important to note that since $\Dr$ itself depends on $\mw$, the equation must be solved via an iterative numerical procedure at each order.
Furthermore, Table~\ref{tab:Wmass} provides an estimate of the impact of these QCD corrections on the effective weak mixing angle, following Eqs.~(\ref{eq:sefflth}) and (\ref{eq:deltakappaQCD}).
The light-quark three-loop contribution to $\delta\mw$ was not previously available in the literature and contributes by an amount comparable to the ultimate experimental precision foreseen at the FCC-ee.
\begin{table}[h]
\centering
\renewcommand{\arraystretch}{1.15}
\begin{tabular}{|c r|r|r|}
\hline
  &  &  \oaas &   \oaasas   \\
\hline\hline
\multirow{2}{*}{$\delta m_W$(MeV)}   & {\footnotesize ${tb}$} & $-59.62$ & $-11.61$ \\
                     & {\footnotesize ${lq}$}  & $-1.98$ & $-0.23$ \\
\hline\hline
\multirow{2}{*}{$10^5 \times \delta \sefff$}   & {\footnotesize ${tb}$}   & $-67.1$ & $-14.3$ \\
                     & {\footnotesize ${lq}$}  & $-3.4$ & $-0.5$ \\
\hline
\end{tabular}
\caption{\oaas\,and \oaasas\, contributions to $\delta \mw$ and $\delta  \sefff$ from heavy ($tb$) and light ($lq$) quark flavors.}
\label{tab:Wmass}
\end{table}


\subsection{Results for $\delta c_Z$ and $\delta c_W$}
\begin{figure}[H]
    \centering
    \includegraphics[height=6cm, width=8.5cm]{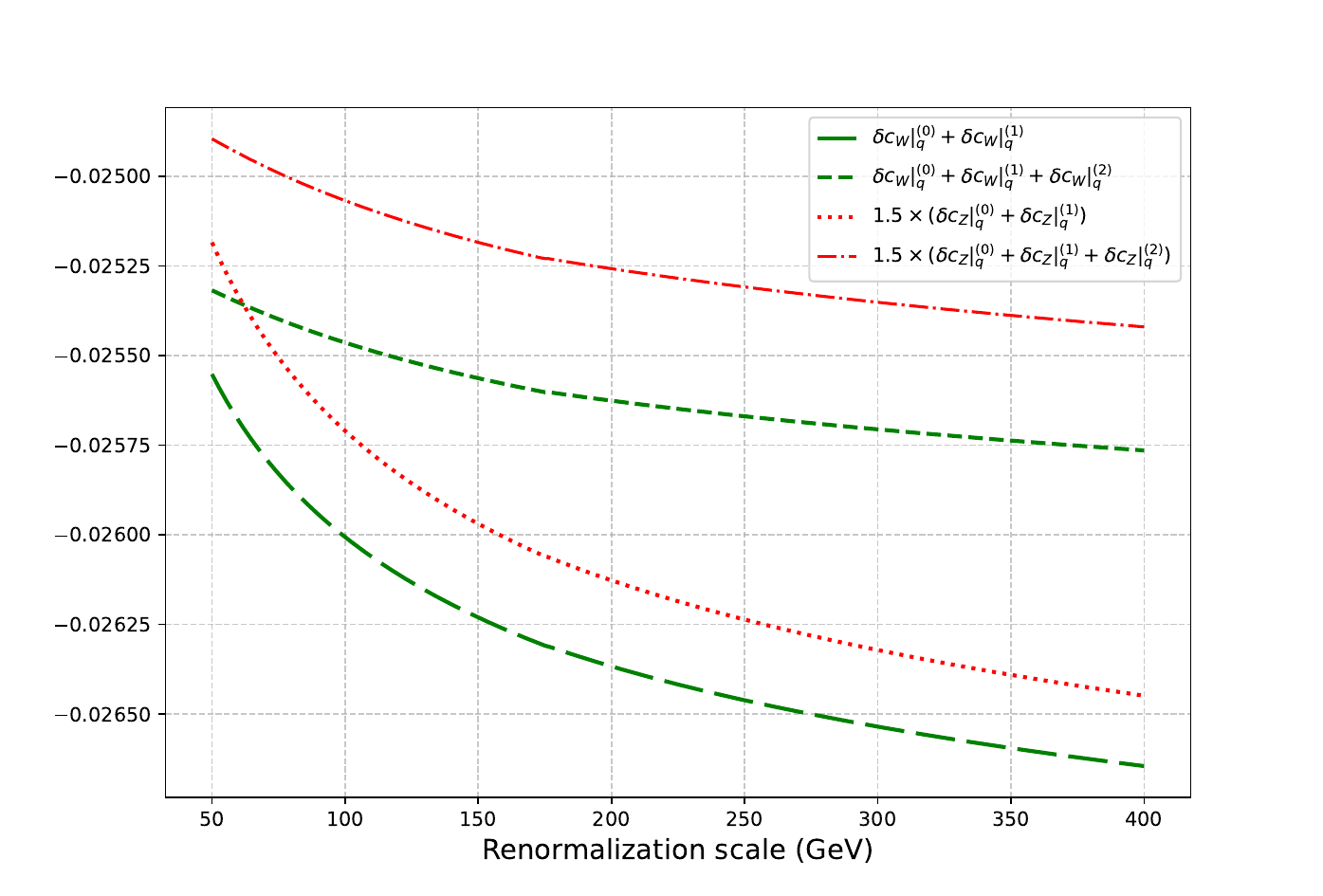}
    \caption{Renormalization scale variation of $\delta c_Z$ and $\delta c_W$ 
     including contributions at different orders.}
    \label{fig:deltacV}
\end{figure}

\noindent
The QCD corrections to the EW first order expressions of the 
UV-finite renomalization combinations $\delta c_V$ affect only the quark self-energies.
They admit a perturbative expansion in $\alpha_s$ as
\begin{align}
 \delta c_V|_q &= - \frac{3 G_F m_t^2}{8 \sqrt{2} \pi^2} \, \frac{1}{s_w^2} \, \bigg\{
  \sum_{m=0}^{\infty} \asr^m \delta c_V^{(m)}  \bigg\} \,.
\end{align}
These contributions are finite and play a crucial role in any process involving the production of OS vector bosons. The semi-analytic expressions for the three-loop corrections, $\delta c_Z^{(2)}$ and $\delta c_W^{(2)}$, are included in the ancillary file. 
Table~\ref{tab:cZW} provides numerical values for these constants using our chosen parameter set
and $\mu_R^2=\mz^2$.
\begin{table}[h]
\centering
\begin{tabular}{|c r|r|r|}
\hline
 $m$ &               &  $\text{Re}(\delta c_Z^{(m)})$~ &   $\text{Re}(\delta c_W^{(m)})$~   \\
\hline\hline
\multirow{2}{*}{0}   & {\footnotesize ${tb}$}  & $ 0.6861$ & $ 0.9448$ \\
                     & {\footnotesize ${lq}$}  & $-0.1898$ & $-0.2083$ \\
\hline
\multirow{2}{*}{1}   & {\footnotesize ${tb}$}  & $-7.4104$ & $-9.7635$ \\
                     & {\footnotesize ${lq}$}  & $-0.7591$ & $-0.8332$ \\
\hline
\multirow{2}{*}{2}   & {\footnotesize ${tb}$}  & $-108.6109$ & $-137.1795$ \\
                     & {\footnotesize ${lq}$}  & $-5.1474$ & $-6.6159$ \\
\hline
\end{tabular}
\caption{Contributions to $\delta c_Z$ and $\delta c_W$ up to \oaasas.}
\label{tab:cZW}
\end{table}
In Fig.~\ref{fig:deltacV}, we illustrate the $\mu_R$ variation of $\delta c_Z$ 
and $\delta c_W$. The inclusion of \oaasas\, contributions reduce the $\mu_R$ dependence.

\section{Conclusion}
\noindent
Precision physics stands at the forefront of the modern particle physics program.
Precision measurements of EW radiative parameters are essential for probing the limits of the SM. To meet the requirements of future electron-positron colliders, the theoretical accuracy must reach a comparable level.

In this work, we utilized advanced perturbative techniques to reevaluate the three-loop QCD vacuum polarization functions for EW gauge bosons.
We have reported, for the first time, the \oaasas\, contributions to the running of the 
$\msbar$-renormalized electric charge.
We further applied the results of the vacuum polarization functions to determine 
the corresponding corrections to $\Drho$, $\Dr$ and $\Dk$.
We have provided a compact analytic expression for $\Drho^{(2)}$ and found full agreement with existing literature.
We have presented $\Delta r^{(2)}$ and $\Delta \kappa^{(2)}$, as a Taylor series in $z_t$ up to ${\mathcal{O}}(z_t^{15})$, and we have included novel \oaasas\, contributions from massless quarks.
We show that these contributions are numerically significant and indispensable for the high-precision determination of $\mw$ and $\sefff$ at the future lepton colliders.
Furthermore, we have presented new results for the UV-finite renormalization combinations 
$\delta c_Z^{(2)}$ and $\delta c_W^{(2)}$.

To facilitate the use of these results in future phenomenological studies, 
we have provided the expressions in an accompanying ancillary file. 
These results represent a necessary step toward the sub-permille theoretical 
accuracy required for the next generation of precision experiments.

\section*{Acknowledgment}
We would like to thank T. Armadillo for cross-checking two-loop results.
We also thank S. Moch and V. Ravindran for their insightful comments on the manuscript. 
N.R. is partially supported by the SERB-SRG under Grant No. SRG/2023/000591.

\biboptions{sort&compress}
\bibliographystyle{apsrev4-1} 
\bibliography{main}

\end{document}